\documentclass[fleqn,twoside]{article}
\usepackage{espcrc2}

\usepackage{epsfig}
\usepackage{wrapfig}                 
                                           
\def\Journal#1#2#3#4{{#1} {\bf #2} (#3) #4}

\def\NPB{{\em Nucl. Phys.}   {\bf B}}
\def\PLB{{\em Phys. Lett.}   {\bf B}}

\def\PRD{{\em Phys. Rev.}    {\bf D}}
\def\ZPC{{\em Z. Phys.}      {\bf C}}
\def\EJC{{\em Eur. Phys. J.} {\bf C}}

\newcommand{\pom}{{I\!\!P}}

\newcommand{\lapprox}{\stackrel{<}{_{\sim}}}
\newcommand{\gapprox}{\stackrel{>}{_{\sim}}}

\title{Inclusive Diffraction at HERA}

\author{F.-P. Schilling\address[desy]
{DESY, Notkestr. 85, D-22603 Hamburg, Germany}%
        \thanks{e-mail address: fpschill@mail.desy.de}
(on behalf of the H1 and ZEUS collaborations)
\thanks{Talk presented at 31st Intl. Conference on 
High Energy Physics ICHEP 2002, Amsterdam}
}
       
\begin{document}

\begin{abstract}
  
  New precision measurements of inclusive diffractive deep-inelastic
  $ep$ scattering interactions, performed by the H1 and ZEUS
  collaborations at the HERA collider, are discussed.  A new set of
  diffractive parton distributions, determined from recent high
  precision H1 data, is presented.

\vspace{1pc}
\end{abstract}

\maketitle

\section{INTRODUCTION}

One of the biggest challenges in our understanding of QCD is the
nature of colour singlet exchange or {\em diffractive} interactions.
The electron-proton collider HERA is an ideal place to study hard
diffractive processes in deep-inelastic $ep$ scattering (DIS).  In
such interactions, the point-like virtual photon probes the structure
of colour singlet exchange, similarly to inclusive DIS probing proton
structure.

\begin{wrapfigure}{L}{0.5\linewidth}      
\centering
\epsfig{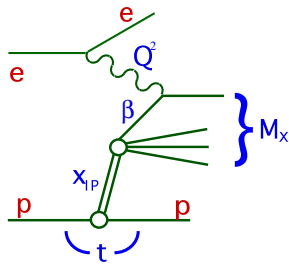}
\caption{Illustration of a diffractive DIS event.}
\label{feyn}
\end{wrapfigure}

At HERA, around $10\%$ of low $x$ events are diffractive
\cite{obsdiff}.  Experimentally, such events are identified by either
tagging the elastically scattered proton in {\em Roman pot}
spectrometers $60-100\rm\ m$ downstream from the interaction point or
by asking for a large rapidity gap without particle production between
the central hadronic system and the proton beam direction.

A diagram of diffractive DIS is shown in Fig.~\ref{feyn}.  A virtual
photon coupling to the beam electron interacts diffractively with the
proton through the exchange of a colour singlet and produces a
hadronic system $X$ with mass $M_X$ in the final state.  If the
4-momenta of the incoming (outgoing) electron and proton are labeled
$l$ ($l'$) and $p$ ($p'$) respectively, the following kinematic
variables can be defined: $Q^2=-q^2=-(l-l')^2$, the photon virtuality;
$\beta = Q^2/q.(p-p')$, the longitudinal momentum fraction of the
struck quark relative to the diffractive exchange; $x_\pom=q.(p-p') /
q.p$, the fractional proton momentum taken by the diffractive exchange
and $t=(p-p')^2$, the 4-momentum squared transferred at the proton
vertex.  Bjorken-x is given by $x=x_\pom \beta$.  For the measurements
presented here typical values of $x_\pom$ are $<0.05$.  $y=Q^2/sx$
denotes the inelasticity, where $s$ is the $ep$ CMS energy.

A diffractive {\em reduced cross section} $\sigma_r^{D(4)}$ can be 
defined via
\begin{eqnarray}
\frac{d^4\sigma^{ep\rightarrow eXp}}{dx_\pom \ dt \ d\beta \ dQ^2}=  
\nonumber \\
\frac{4\pi\alpha^2}{\beta Q^4}\left(1-y+\frac{y^2}
{2}\right)\sigma_r^{D(4)}(x_\pom,t,\beta,Q^2) \ , 
\end{eqnarray}
which is related to the diffractive structure
functions $F_2^D$ and the longitudinal $F_L^D$ by 
\begin{equation}
\sigma_r^D = F_2^D-\frac{y^2}{2(1-y+\frac{y^2}{2})} F_L^D \ . 
\end{equation}
Except at the highest $y$, $\sigma_r^D=F_2^D$ to a very good
approximation. If the outgoing proton is not detected, the
measurements are integrated over $t$: $\sigma_r^{D(3)} = \int {\rm d}t
\ \sigma_r^{D(4)}$.

\section{FACTORIZATION PROPERTIES OF DIFFRACTIVE DIS}

The proof \cite{collins} that QCD hard scattering factorization is
valid for diffractive DIS justifies the expression of
$\sigma_r^{D(4)}$, at fixed $x_\pom$ and $t$, as a convolution of
diffractive proton parton distributions (dpdf's) $f_i^D$ and partonic
cross sections $\hat{\sigma}^{\gamma^*i}$:
\begin{eqnarray}
\sigma_r^{D(4)} \sim \sum_i 
\hat{\sigma}^{\gamma^*i}(x.Q^2) \otimes
f_i^D(x,Q^2,x_\pom,t) \ .
\label{equ:diffpdf}
\end{eqnarray}
The $f_i^D$ should obey the DGLAP evolution equations and the
$\hat{\sigma}^{\gamma^* i}$ are the same as for standard DIS.  In
consequence, the framework of NLO QCD can be applied to diffractive
DIS in a similar way as to inclusive DIS.

An extra assumption which is often made is that the $(x_\pom,t)$
dependence of $\sigma_r^D$ can be factored out (Regge factorization)
into a {\em Pomeron flux factor} $f_\pom(x_\pom,t)$:
\begin{equation}
f_i^D(x_\pom,t,\beta,Q^2) = f_\pom(x_\pom,t) \ f_i^\pom(\beta,Q^2) \ ,
\label{eq:reggefact}
\end{equation}
which is often parameterised using Regge phenomenology as
\begin{equation}
 f_\pom(x_\pom,t) = x_\pom^{1-2\alpha_\pom(t)} e^{bt} \ , 
\label{eq:reggeflux}
\end{equation} 
where $\alpha_\pom(t)=\alpha_\pom(0)+\alpha_\pom't$ is the pomeron
trajectory with intercept $\alpha_\pom(0)$. The $f_i^\pom(\beta,Q^2)$
are {\em Pomeron pdf's}.  Although there is no firm basis in QCD for
Eq.~\ref{eq:reggefact}, it is approximately consistent with present
data.

\section{EXPERIMENTAL RESULTS}

\begin{figure}[bt]
\centering
\epsfig{file=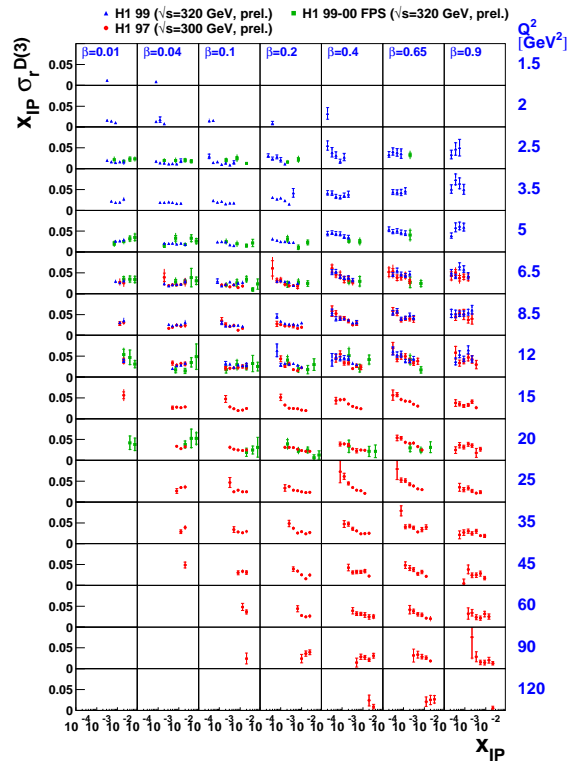,width=\linewidth}
\caption{Compilation of new measurements of the diffractive 
reduced cross section
$\sigma_r^D$ by H1.}
\label{fig:stamph1}
\end{figure}

\begin{figure}[bt]
\centering
\epsfig{file=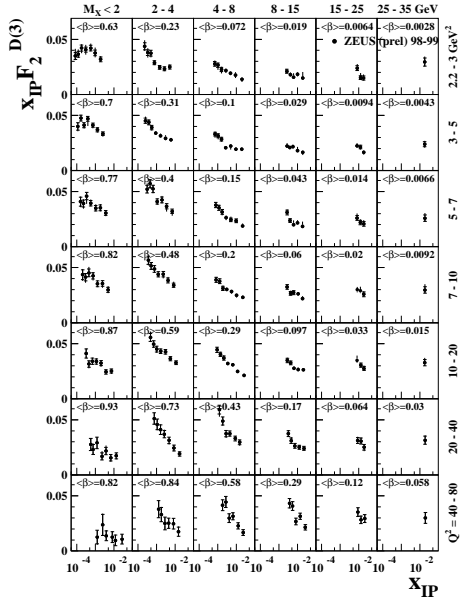,width=0.8\linewidth}
\caption{New measurement of the structure function $F_2^{D(3)}$
by ZEUS.}
\label{fig:stampzeus}
\end{figure}

\begin{figure}[bt]
\centering
\epsfig{file=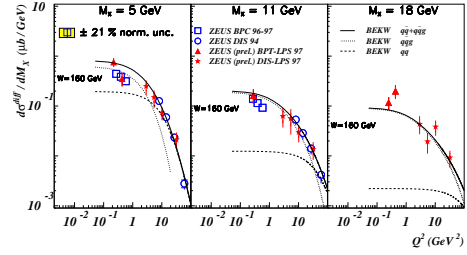,width=0.8\linewidth}
\caption{New measurement of the diffractive cross section at low $Q^2$ 
by ZEUS.}
\label{fig:zeusloq2}
\end{figure}

\subsection{The New Datasets}

The H1 collaboration has performed several new measurements of
diffractive DIS, presented in the form of the reduced cross section
$\sigma_r^{D(3)}$, which are summarised in Fig.~\ref{fig:stamph1}. Two
high precision measurements in the kinematic range $1.5<Q^2<120 \rm\ 
GeV^2$ (labeled `H1 99' \cite{ichepf2d99mb} and `H1 97'
\cite{ichepf2d97}) are based on the rapidity gap method.  With respect
to previous H1 data \cite{h1f2d94}, a factor of typically 5 more data
were analysed and the kinematic range was extended significantly to
lower $Q^2$ and $\beta$. Another new measurement using a forward
proton spectrometer covers $2.5<Q^2<20 \rm\ GeV^2$ (`H1 99-00 FPS',
\cite{ichepfps}).  Good agreement between the different data sets and
measurement techniques is observed.

The ZEUS collaboration has presented new data \cite{zeusfpc} on
$F_2^{D(3)}$ for $2.2<Q^2<80 \rm\ GeV^2$, shown in
Fig.~\ref{fig:stampzeus}.  Diffractive events are selected by
reconstructing a low value of $M_X$ in the detector.  The data
correspond to a factor 2 increase in integrated luminosity with
respect to previous results \cite{ZEUS:94}. ZEUS has also measured
(Fig.~\ref{fig:zeusloq2}, \cite{zeusloq2}) the diffractive cross
section in the transition region $0.03<Q^2<0.6 \rm\ GeV^2$ between DIS
and quasi-real photon-proton interactions ($Q^2\sim0$). Also shown in
Fig.~\ref{fig:zeusloq2} is a fit based on a dipole model \cite{bekw}
in which the photon fluctuates into $q\bar{q}$ and $q\bar{q}g$ states.

\subsection{$t$ and $\Phi_{ep}$ Dependences}

\begin{figure}[bt]
\centering
\epsfig{file=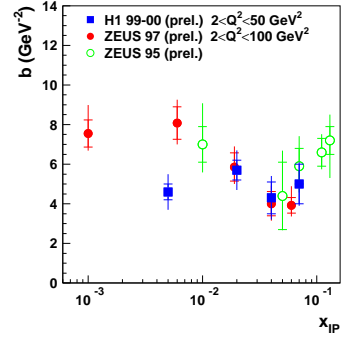,width=0.6\linewidth}
\caption{The slope parameter $b$ from fits to 
$\frac{d\sigma}{dt} \sim e^{bt}$ at different values of $x_\pom$.
}
\label{fig:bslopes}
\end{figure}

Both collaborations have measured the $t$ dependence of the cross
section using their forward proton spectrometers. In the measured
range $-0.45 \lapprox t<-|t|_{min}$, an exponential dependence
$\frac{d\sigma}{dt}\sim e^{bt}$ is observed and the slope parameter
$b$ is determined for different values of $x_\pom$
(Fig.~\ref{fig:bslopes}).  Models based on Regge phenomenology predict
an increasing steepness of the $t$ dependence with energy ({\em
  shrinkage}): $b=b_0+2\alpha'\log{\frac{1}{x_\pom}}$.  As can be seen
in Fig.~\ref{fig:bslopes}, the data at low $x_\pom$ are so far
inconclusive.

ZEUS has studied \cite{zeuslps} the cross section dependence on the
azimuthal angle $\Phi_{ep}$ between the electron and proton scattering
planes, which is sensitive to the longitudinal cross section
$\sigma_L^D$ via the interference between transverse and longitudinal
photon induced contributions, which introduces an asymmetry in
$\Phi_{ep}$. The measured asymmetries are small and, within the
present statistical uncertainties, compatible with zero.

\subsection{$x_\pom$ Dependence}

\begin{figure}[bt]
\centering
\epsfig{file=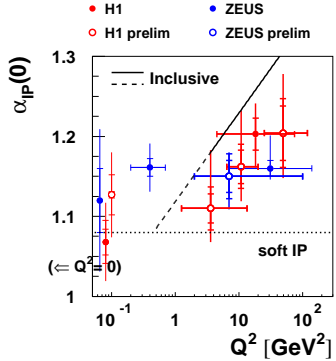,width=0.59\linewidth}
\caption{Measurements of the effective pomeron intercept $\alpha_\pom(0)$
for different values of $Q^2$, compared with a parameterisation from
inclusive DIS.}
\label{fig:alpha}
\end{figure}

By making fits to the $x_\pom$ dependence of the diffractive DIS data
using Eqs.~\ref{eq:reggefact},\ref{eq:reggeflux}, values for the
effective pomeron intercept $\alpha_\pom(0)$ are determined.  Results
for measurements at different $Q^2$ are shown in Fig.~\ref{fig:alpha}
and compared with the {\em soft pomeron} value of 1.08 as well as with
a parameterisation based on inclusive DIS data, obtained from fits of
the form $F_2(x,Q^2)\sim c x^{-\lambda(Q^2)}$ where
$\lambda=\alpha_\pom(0)-1$.  The diffractive data suggest an increase
of $\alpha_\pom(0)$ with $Q^2$ and are significantly higher than the
soft pomeron value for $Q^2\gapprox 10 \rm\ GeV$. However, the naive
expectation that the diffractive cross section grows twice as fast
with energy as the inclusive one, corresponding to
$\alpha^{dif.}_\pom(0)=\alpha^{incl.}_\pom(0)$, is hard to reconcile
with the present data at high $Q^2$.

\subsection{Ratio of Diffractive to Inclusive Cross Section}

\begin{figure}[bt]
\centering
\epsfig{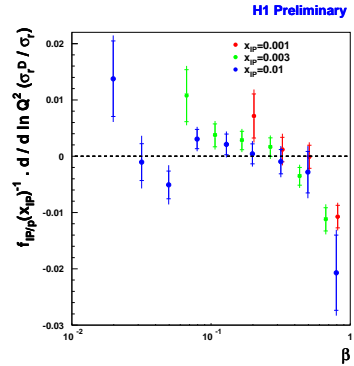}
\caption{Logarithmic $Q^2$ derivatives of the ratio $\sigma_r^D/\sigma_r$ 
from H1, divided by $f_\pom(x_\pom)$, shown as a function of 
$\beta$ for different $x_\pom$.}
\label{fig:ratio}
\end{figure}

H1 has studied \cite{ichepf2d97} the logarithmic $Q^2$ dependence of
the ratio of the diffractive to the inclusive DIS reduced cross
sections, presented as the quantity $f_\pom(x_\pom)^{-1} \ {\rm d} /
{\rm d} {\rm ln} Q^2 (\sigma_r^D / \sigma_r)$ in Fig.~\ref{fig:ratio}
for different $\beta$ and $x_\pom$ values. Taking out the factorising
$x_\pom$ dependence, the same behaviour is observed at different
$x_\pom$: At low $\beta$, the logarithmic $Q^2$ dependences of
$\sigma_r^D$ and $\sigma_r$ are very similar, indicating that the
ratio of the diffractive to the total proton gluon density is
approximately constant in this region.  By contrast, at the highest
$\beta \gapprox 0.5$, the logarithmic $Q^2$ derivative becomes
negative, suggestive of the presence of $Q^2$-suppressed higher twist
contributions to $\sigma_r^D$ for $\beta\rightarrow 1$. However, this
behaviour can also be explained by DGLAP evolution due to the the
kinematic limit $x=x_\pom$ ($\beta=1$) for gluon radiation in the case
of $\sigma_r^D$.

\section{$\beta$ AND $Q^2$ DEPENDENCES OF $\sigma_r^D$ AND
NLO QCD FIT TO H1 DATA}

\begin{figure}[bt]
\centering
\epsfig{file=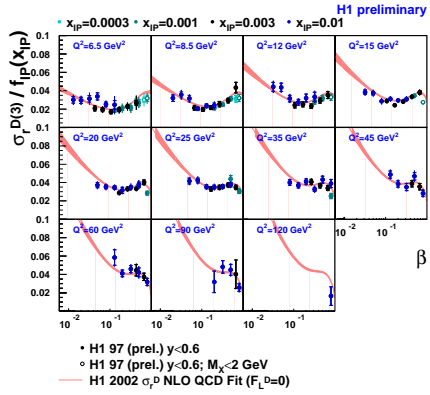,width=0.74\linewidth}
\caption{$\beta$ dependence of $\sigma_r^D$ from H1, 
scaled by $f_\pom(x_\pom)^{-1}$ and compared with the NLO QCD fit.
}
\label{fig:betadep.flux}
\end{figure}

\begin{figure}[bt]
\centering
\epsfig{file=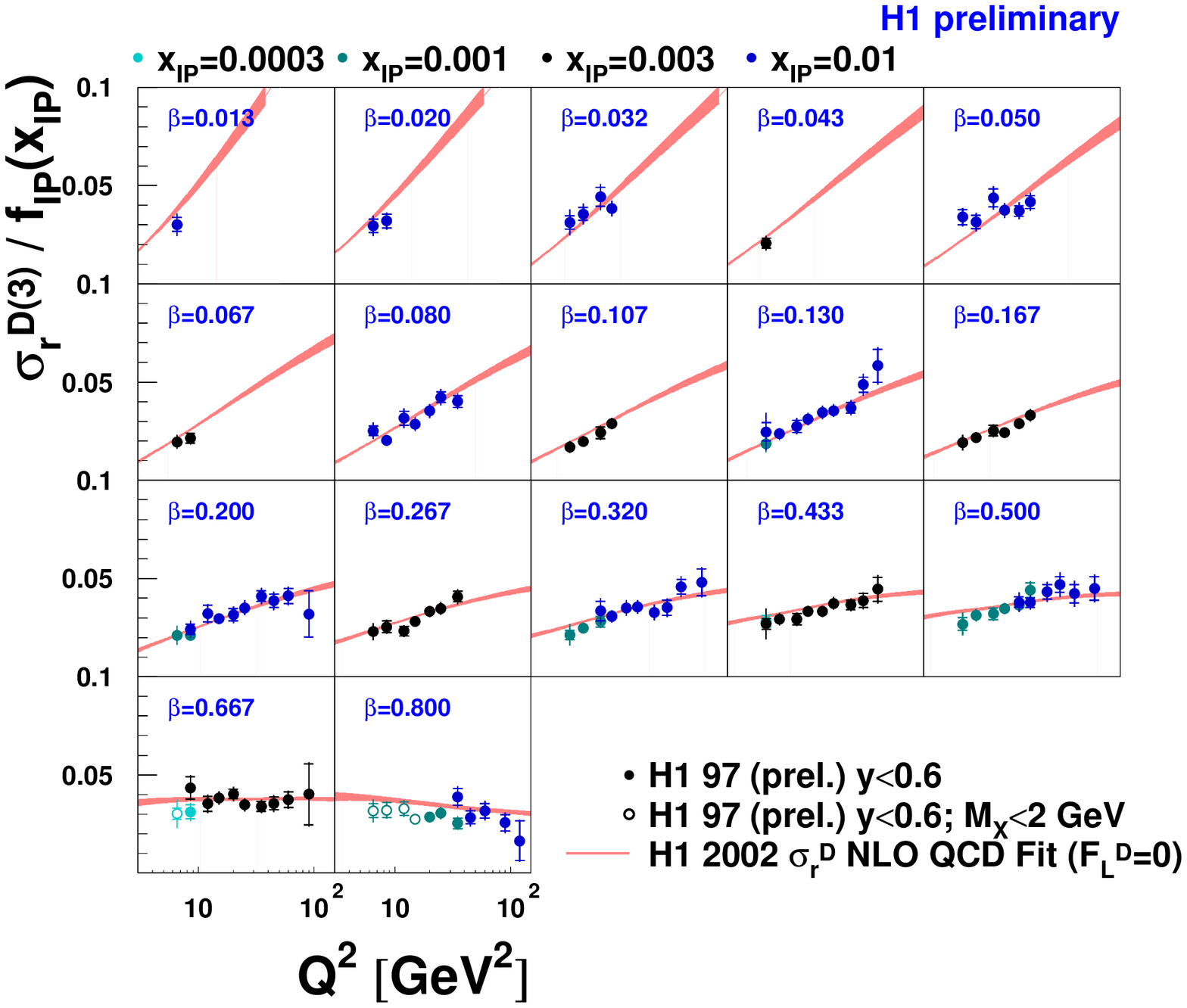,width=0.8\linewidth}
\caption{$Q^2$ dependence of $\sigma_r^D$ from H1, 
scaled by $f_\pom(x_\pom)^{-1}$ and compared with the NLO QCD fit.
}
\label{fig:q2dep.flux}
\end{figure}

\begin{figure}[!bt]
\centering
\epsfig{file=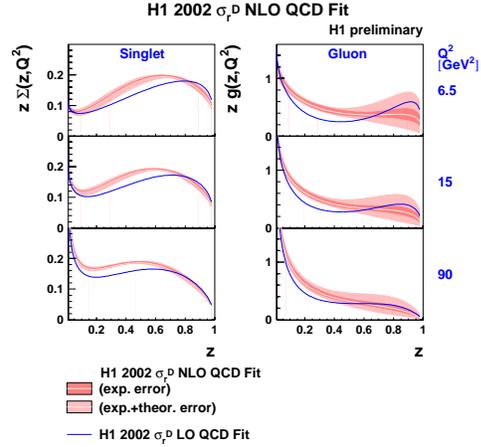,width=0.825\linewidth}
\caption{Diffractive singlet (left) and gluon (right)
distributions from the H1 NLO QCD fit.
}
\label{fig:h1pdfs}
\end{figure}

H1 has presented \cite{ichepf2d97} a new NLO DGLAP QCD fit to the
diffractive reduced cross section data, from which diffractive parton
densities are determined. The shape of the dpdf's is assumed to be
independent of $x_\pom$ (Regge factorization), in agreement with the
data, as can be seen in Figs.~\ref{fig:betadep.flux} and
\ref{fig:q2dep.flux}. The precisely measured $\beta$ and $Q^2$
dependences of the data are compared with the fit result, dividing out
the $x_\pom$ dependence, parameterised as stated in
Eq.~\ref{eq:reggeflux}.

The diffractive exchange is parameterised by a light flavour singlet
and a gluon distribution at a starting scale $Q_0^2=3 \rm\ GeV^2$ and
evolved to higher $Q^2$ using the standard NLO DGLAP equations.  For
the first time in diffraction, the experimental and model
uncertainties are propagated to obtain error bands for the dpdf's.

The result of the fit is presented in Fig.~\ref{fig:h1pdfs}.  The
diffractive pdf's remain large up to large fractional momenta $z$ (or
$\beta$) and are dominated by the gluon distribution. In total $75\%$
of the exchange momentum is carried by gluons at $Q^2=10\rm\ GeV^2$.
To test factorization, the dpdf's can be used for updated comparisons
with diffractive final state data from HERA and the TEVATRON
\cite{ichepf2d97,savin}.

\section{CONCLUSIONS}

Several new high precision data sets of inclusive diffractive DIS have
recently become available. In the case of the rapidity gap method, the
data are now systematically limited at low $Q^2$.  These data are very
useful to constrain the free parameters of QCD motivated models (e.g.
dipole models) for diffractive DIS and to distinguish between
different approaches.  In particular, they have been used for the
determination of a new generation of diffractive parton distributions
-- including their uncertainties -- in the framework of NLO QCD,
enabling tests of QCD factorization as applied to diffraction by
comparisons with final state cross sections.

\subsection*{Acknowledgements}

I thank my colleagues from H1 and ZEUS for their 
work reflected in this contribution.

\end{document}